\documentclass[aps,pra,superscriptaddress,twocolumn,showpacs]{revtex4-1}
\usepackage{epsfig}
\usepackage{graphicx}
\usepackage{amsmath}
\usepackage{comment}
\usepackage{braket}
\usepackage{bm}

\allowdisplaybreaks

\newcommand{\Paris}{Universit\'e Sorbonne Paris Nord, Laboratoire de Physique des Lasers, F-93430, Villetaneuse, France}
\newcommand{\CNRS}{CNRS, UMR 7538, LPL, F-93430, Villetaneuse, France}
\newcommand{\JILA}{JILA, NIST and Department of Physics, University of Colorado, Boulder, USA}
\newcommand{\CTQM}{Center for Theory of Quantum Matter, University of Colorado, Boulder, CO 80309, USA}
\newcommand{\QuICS}{Joint Center for Quantum Information and Computer Science, NIST and University of Maryland, College
Park, Maryland 20742, USA}
\newcommand{\JQI}{Joint Quantum Institute, NIST and University of Maryland, College Park, Maryland 20742, USA}
\newcommand{\Lyon}{Univ Lyon, Ens de Lyon, CNRS, Laboratoire de Physique, F-69342 Lyon, France}

\begin{document}

\DeclareGraphicsExtensions{.eps,.EPS,.pdf}

\title{ Measuring bipartite spin correlations of lattice-trapped dipolar atoms}

\author{Youssef Aziz Alaoui}\thanks{These two authors contributed equally}
\affiliation{\Paris}
\affiliation{\CNRS}
\author{Sean R. Muleady}\thanks{These two authors contributed equally}
\affiliation{\JILA}
\affiliation{\CTQM}
\affiliation{\QuICS}
\affiliation{\JQI}
\author{Edwin Chaparro}
\affiliation{\JILA}
\affiliation{\CTQM}
\author{Youssef Trifa}
\affiliation{\Lyon}
\author{Ana Maria Rey}
\affiliation{\JILA}
\affiliation{\CTQM}
\author{Tommaso Roscilde}
\affiliation{\Lyon}
\author{Bruno Laburthe-Tolra}
\affiliation{\CNRS}
\affiliation{\Paris}
\author{Laurent Vernac}
\affiliation{\Paris}
\affiliation{\CNRS}

\begin{abstract}

We demonstrate a bipartition technique using a super-lattice architecture to access correlations between alternating planes of a mesoscopic array of spin-3 chromium atoms trapped in a 3D optical lattice.
Using this method, we observe that out-of-equilibrium dynamics driven by long-range dipolar interactions lead to spin anti-correlations between the two spatially separated subsystems.
Our bipartite measurements reveal a subtle interplay between the anisotropy of the 3D dipolar interactions and that of the lattice structure, without requiring single-site addressing.
We compare our results to theoretical predictions based on a truncated cumulant expansion and a new cluster semi-classical method that we use to investigate correlations at the microscopic scale. Comparison with a high-temperature analytical model reveals quantum thermalization at a high negative spin temperature.

\end{abstract}
\date{\today}
\maketitle

{\it Introduction.}
The study of fluctuations between sub-ensembles of a quantum system is crucial to reveal some of the most fundamental properties of quantum mechanics~\cite{EPR1935}. It underlies concepts such as entanglement~\cite{Horodecki2009}, EPR steering~\cite{Cavalcanti2016}, and Bell correlations~\cite{Brunner2014}, and plays an essential role in quantum scrambling~\cite{Hosur2016,Styliaris2021} and quantum thermalization~\cite{Poilblanc2011,Dalessio2016,Morietal2018}. When individual addressing of each particle is available, advanced techniques have been developed to demonstrate the measurement of entanglement entropy \cite{Islam2015,Brydges2019,Nikman2021}, purity certification \cite{Kaufman2016}, quantum thermalization \cite{Neill2016,Kaufman2016,Christakis2023}, high-fidelity quantum gates \cite{Evered2023}, quantum scrambling behavior \cite{Blok2021}, and measurement of out-of-time-order correlators \cite{Li2017, Garttner2017}. For mesoscopic systems, for which full quantum tomography is highly impractical, demonstration of EPR correlations was obtained by measuring the relative fluctuations between two subsystems of a bulk BEC in the single-mode regime\cite{Lange2018,Fadel2018, Kunkel2018}.

Platforms harnessing dipolar interacting particles are of particular interest for studying the propagation of quantum correlations \cite{Kuwahara2021, Lerose2020,Eisert2013,Perlin2020,Comparin2022,Roscilde2023,Trifa2023,Gong2017}, which can be very different compared to those induced by finite-range interactions, due to the high group velocity of elementary excitations \cite{Frerot2018,Cevolani2018,Chen2023}. In these platforms, the 3D anisotropy plays an unavoidable role \cite{Lepoutre2019,Patscheider2020}.
Experimentally, individual addressing in dipolar systems is mostly available for particles with strong electric dipole moment, such as in Rydberg atoms \cite{Browaeys2020}; heteronuclear molecules in optical lattices \cite{Yan2013,Christakis2023}; or tweezer arrays \cite{Kaufman2021,Bao2023,Holland2023}.
 In the case of magnetic atoms \cite{Chomaz2023}, which are the focus of this study, the use of short-period lattices, necessary to boost the strength of the interactions, makes individual addressing challenging  (see however a recent realization in 2D \cite{Su2023}). Moreover, the use of a quantum gas microscope in 3D remains extremely challenging \cite{Legrand2024}. It is thus relevant to develop new tools based on collective measurements of subsystems, which allow one to tackle the correlations and entanglement in dipolar 3D systems without the need for individual addressing.

Here we implement bipartition within a 3D optical lattice loaded with strongly magnetic chromium atoms. We use a super-lattice in order to separate even and odd planes along a particular lattice vector, and thus perform bipartite measurements that ensure a large interface between both subsystems. We measure the growth of anisotropic spin correlations at the level of the standard quantum projection noise, and demonstrate anti-correlations between the collective magnetization of the two subsystems. We also compare our results to different advanced numerical methods.  In particular, we introduce a useful refinement of the generalized discrete
truncated Wigner semi-classical approximation (GDTWA) \cite{Zhu2019}), in which certain local correlations are treated exactly. Furthermore, we perform calculations for the thermalized state that is expected at long time \cite{Lepoutre2019}, using a high-temperature expansion.
Our combined experimental and theoretical analysis demonstrates that bipartite correlations are inherently sensitive to the anisotropy of the system, and to the value and sign of the effective spin temperature at equilibrium.

\begin{figure}
\centering
\includegraphics[width= 3.3 in]{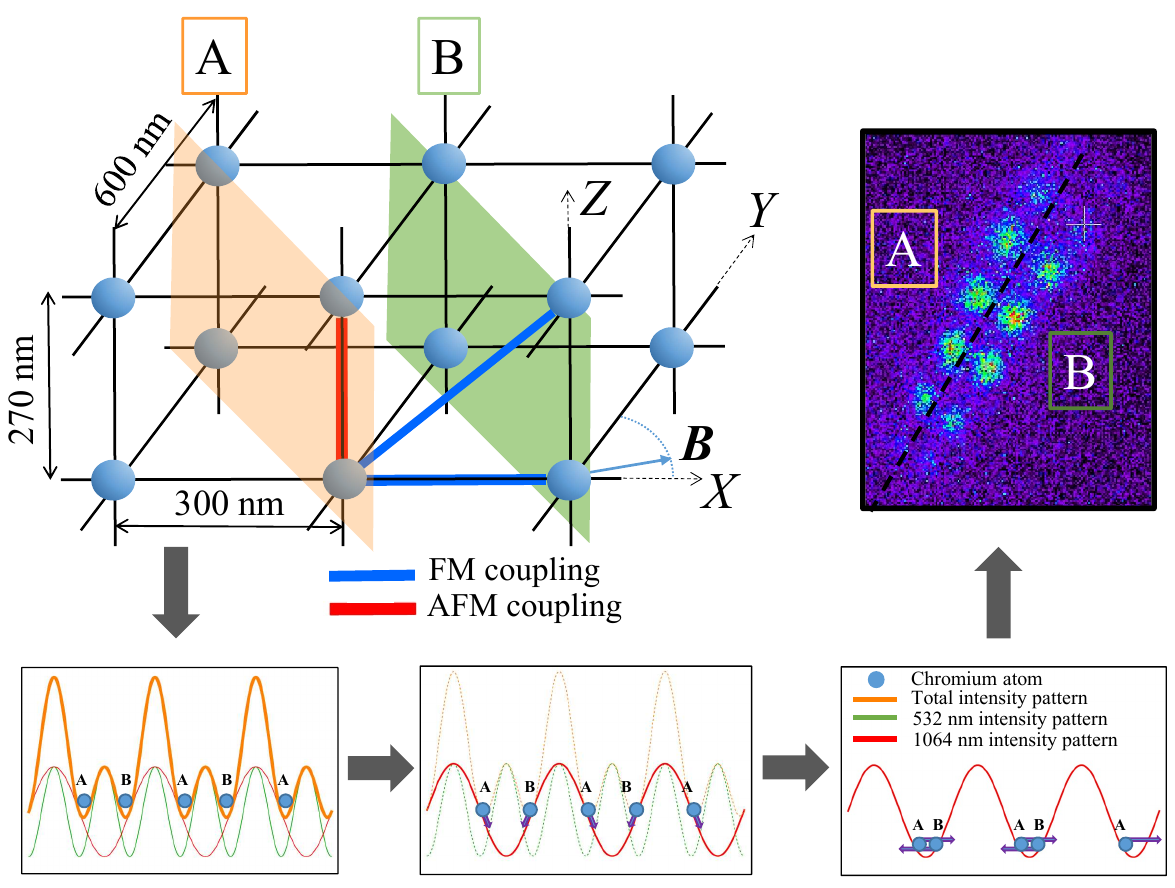}
\caption{\setlength{\baselineskip}{6pt} {\protect\scriptsize Sketch of the experiment, with gray arrows indicating the steps of our bipartition sequence. The spin dynamics take place in a $532$ nm 3D anisotropic lattice, which is dominated by a combination of ferromagnetic (FM) and anti-ferromagnetic (AFM) dipolar couplings in the $XZ$ plane. The addition of a infrared (IR) laser at the end of the dynamics creates a double-well structure that gives rise to two spin subsystems $A$ and $B$. Atoms then evolve in the pure IR lattice for a quarter period, which allows for spatial separation of the two subsystems after time of flight as a result of opposite momenta conveyed to the atoms in the IR lattice. Combining bipartition to Stern-Gerlach separation, measurement of the magnetization of both subsystems is obtained with fluorescence imaging.}}
\label{fig1}
\end{figure}

{\it  Description of the spin system and bipartition protocol.}
The spin $s=3$ chromium atoms are pinned at the antinodes of a 3D lattice. The presence of an external $0.75$ gauss magnetic field \textbf{B}, which is strong enough to generate Zeeman splittings larger than nearest-neighbor dipolar interactions, ensures that only  processes that conserve the total magnetization are energetically allowed. The dynamics are thus described by the following  effective xxz (\textbf{B} $\parallel$ \textbf{z}) Hamiltonian, together with a one-body term:
\begin{equation}
\begin{split}
\hat H=\sum_{i> j}^{N} V_{i,j} \left[ \hat s_{z,i} \hat s_{z,j} -\frac{1}{2} \left( \hat s_{x,i} \hat s_{x,j} + \hat s_{y,i} \hat s_{y,j} \right) \right]\\+ B_{\rm Q}\sum_i^N \hat s_{z,i}^2
\end{split}
\label{secular}
\end{equation}
with $V_{i,j}=V_{dd} \left( 1-3 \cos ^2 \theta _{i,j}\right)/r_{i,j}^3$, $V_{dd}= \mu_0 (g_L \mu_B)^2/4 \pi$, where $\mu_0$ is the magnetic permeability of vacuum, $g_L \simeq 2$ the Land\'e factor, and $\mu_B$ the Bohr magneton. $r_{i,j}$ is the distance between atoms, and $\theta_{i,j}$ the angle between their inter-atomic axis and the external magnetic field. ${\bf{\hat s}}_i=(\hat s_{x,i},\hat s_{y,j},\hat s_{z,i} )$ are spin-3 angular momentum operators  for  atom $i$ when a site is populated by a single atom. $B_{\rm Q}$ describes the strength of a quadratic Zeeman term that accounts for tensor light shifts created by the trapping lasers \cite{Lepoutre2019}.
The lattice, implemented with five laser beams at $\lambda_L=532$ nm \cite{SuppMat}, is anisotropic. It corresponds to a 1D array of relatively well separated planes, each containing a 2D lattice ($XZ$ planes in Fig.~\ref{fig1}). We emphasize that the quantization axis $z$, set by \textbf{B}, is nearly parallel to these planes and differs from the spatial $Z$ axis, to which it is perpendicular.

In a time-dependent experiment that starts from a non-correlated spin state, pairwise entanglement due to dipolar interactions is expected to grow most rapidly at short range. To best reveal this growth of quantum correlations in our optical lattice system using bipartition, we overlap the native lattice structure with a super-lattice, creating an array of double-wells \cite{Anderlini2007,Trotzky2008} that defines two interleaved sub-ensembles, which we refer to as $A$ and $B$. Compared to a bipartition defined by two spatially separated ensembles meeting at a single plane, our scheme increases the number of nearest-neighbour links across the partition by a factor of $N^{1/3}$ where $N$ is the total number of particles in the system, similarly enhancing the correlations between the two subsystems.

The bipartition scheme is detailed in Fig.~\ref{fig1}. After dynamically evolving the system in the short period lattice only, we adiabatically superimpose a $1064$ nm retro-reflected beam overlapped with one of the retro-reflected laser beams used for the native lattice. We then abruptly switch off the $532$ nm lattice, and atoms are left to move in the one-dimensional $1064$ nm lattice only, for a duration that corresponds to roughly one-fourth of the oscillation time within one lattice site. This effectively implements a $\pi /2$ rotation in phase space, translating the opposing positions of the atoms into opposite momenta \cite{Asteria2021}. A time of flight of 14 ms then separates the atoms in two subsystems that correspond to the original $A$ and $B$ subsystems. Stern-Gerlach separation during time of flight allows measurement of the seven spin components within each subsystem, for which we use fluorescence imaging onto an Electron Multiplying-CCD camera.

The shape of the double-well structure acutely depends on the relative phase of the interference fringes associated with the two lasers of the bichromatic lattice at the position of the atoms. An appropriate frequency relation between the two lasers ensures that the optical potential takes the form depicted in Fig.~\ref{fig1}, bottom left. A relative frequency drift of the two lasers leads to the appearance of a third spin ensemble when attempting bipartition (see \cite{SuppMat}). Corrections of such a drift requires a change in frequency amplitude of up to $\Delta f\approx 80$ MHz, set by $2\pi \Delta f L/c=\pi/4$, with $L\approx 50$ cm the distance between the atoms and the retro-reflection mirror producing standing waves.

\begin{figure*}
\centering
\includegraphics[width= 7 in]{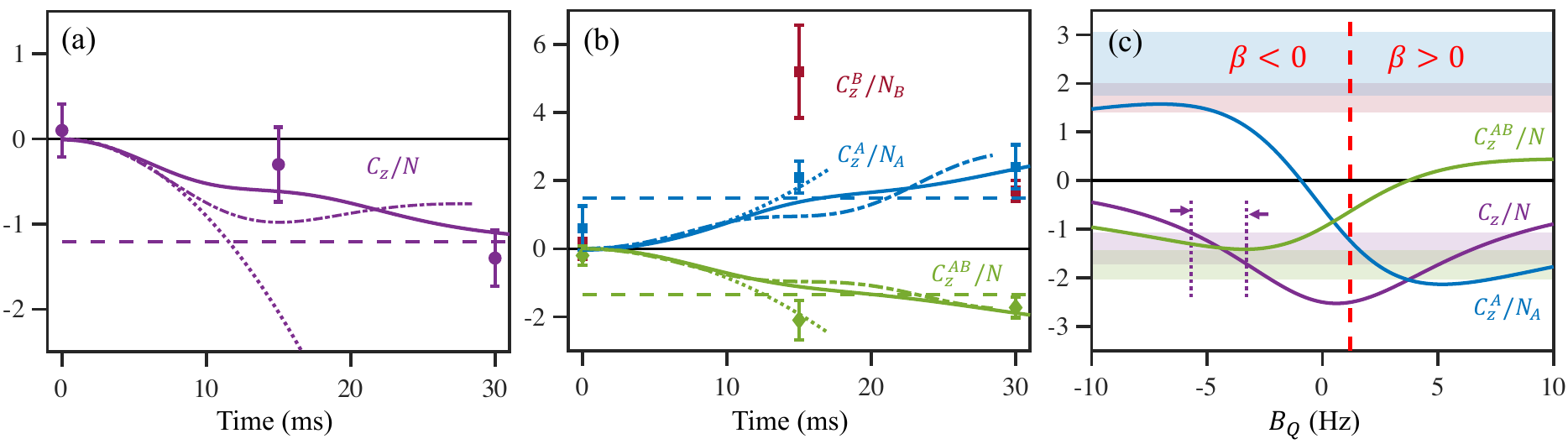}
\caption{\setlength{\baselineskip}{6pt} {\protect\scriptsize Experimental results and model comparisons. For experimental data, the full error bars correspond $\pm1$ statistical standard error. Dynamics of collective (a) and sub-systems (b) correlators: lines show theoretical results obtained via a short time expansion (dotted), truncated cumulant calculations (dot-dashed), cluster-GDTWA simulations (full lines), and a high-temperature expansion (dashed). (c) Values of the correlators in the thermalized state as a function of the quadratic term $B_{\rm Q}$. Solid lines denote the values of $C_z/N$ (purple),$C_z^{A,B}/N_A$ (blue) and $C_z^{AB}/N$ (green). The shaded regions show the corresponding confidence intervals for experimental data at 30 ms, where we also include the confidence interval for $C_z^B/N_B$ as its own separate region (red shaded region).  The dashed red line indicates $B_Q$ for which the state attains an infinite temperature $\beta=0$, with $\beta <0$ ($\beta>0$) to the left (right) of this line. The region bounded by the purple dot-dashed lines denotes the range of $B_Q<0$ compatible with the measured $C_z/N$ errorbars at 30 ms.}}
\label{fig2}
\end{figure*}

{\it Collective fluctuations and pairwise correlators.}
Measurement of the subsystems' collective spin populations allows us to infer the following two-point correlators associated with the magnetization:
\begin{equation}
C_z^{\sigma, \sigma'}=\sum_{i\in {\sigma},j\in {\sigma'},i\neq j}\Big(\left<\hat{s}_{z,i}\hat{s}_{z,j}\right>-\left<\hat{s}_{z,i}\right>\left<\hat{s}_{z,j}\right>\Big).
\label{defcorrel}
\end{equation}
From this, we can define the correlator for the whole sample, which we denote $C_z$ for simplicity ($ \sigma=\sigma'= A\cup B$); the subsystems correlators $C_z^A$ and $C_z^B$ ($\sigma=\sigma'=A(B)$); and the intersystem correlator $C_z^{AB}$ ($\sigma=A, \sigma'= B$).
Defining subsystem spin operators $\hat S_z^\sigma=\sum_{i\in \sigma}\hat s_{z,i}$, we note that $C_z^{AB}={\rm Cov}(\hat{S}_z^A,\hat{S}_z^B)$.

$C_z,C_z^A,C_z^B$ are related to fluctuations of the spin component $\hat S_z^\sigma$ \cite{Mouloudakis2021}: $C_z^{\sigma}=\rm{Var}(\hat S_z^{\sigma})-\Sigma_z^{\sigma}$,
with $\rm{Var}(\hat S_z^{\sigma})$ the variance of $\hat S_z^{\sigma}$, and $\Sigma_z^{\sigma}=\sum_{i\in \sigma}\left<\hat s_{z,i}^2\right>-\left<\hat s_{z,i}\right>^2$ \cite{Mouloudakis2021}.
In our system, we can neglect inhomogeneities \cite{Alaoui2022}, so that we assume $\left<\hat s_{z,i}\right>=0$ $\forall i$, and  $\Sigma_z ^{\sigma}$ can be evaluated from spin populations $P_{m_s}^ {\sigma}$:  $\Sigma_z ^ {\sigma}=\sum_{i\in \sigma}P_{m_s}^{\sigma} m_s^2$ ($-3\leq m_s\leq3)$~\cite{Alaoui2022}.

As the collective spin $\hat{S}_z = \hat{S}_z^A + \hat{S}_z^B$ commutes with the Hamiltonian Eq.~\eqref{secular}, ${\rm{Var}}(\hat S_z)$ is conserved and expected to remain equal to its initial value (in our case $3N/2$ for a spin coherent state following a $\pi/2$ rotation, see below); on the other hand, variances of the subsystem magnetizations are free to vary from their initial values ($3N_{A,B}/2$, for $N_{A,B}$ the total number of particles in each subsystem) as correlations develop between the two subsystems:
\begin{equation}
{\rm{Var}}(\hat S_z)={\rm{Var}}(\hat S_z^A)+{\rm{Var}}(\hat S_z^B)+2 C_z^{AB}
\label{SumVar}
\end{equation}
Therefore, our bipartition scheme allows to probe correlations in each subsystem, assuming homogeneity, as well as correlations between the subsystems, which are quantified by the covariance featured in the right part of Eq.~\eqref{SumVar}, \textit{without} any assumption on homogeneity.

{\it  Experimental results.}
Initially, the lattice contains an average of $1.5 \times 10^4$ spin-3 atoms, polarized in the minimal Zeeman energy state $m_s=-3$. A deep Mott insulator state regime is reached, excluding transport during the duration of dynamics. To initialize the dynamics, atoms are quenched to an excited state
by use of a radio frequency $\pi/2$ pulse. This prepares a coherent spin state with all spins pointing orthogonal to  \textbf{B}. The out-of-equilibrium spin dynamics driven by dipole-dipole interactions lead to an evolution of the number of atoms measured in each of the different Zeeman states along the quantization axis set by \textbf{B}. From measurement of spin populations, we obtain intraparticle variances $\Sigma_z^{\sigma}$ and ensemble variances ${\rm Var}(\hat S_z^{\sigma})$, and infer $C_z,C_z^A,C_z^B$ for three varying durations of spin dynamics. The intersystem correlators $Cz^{AB}$ are obtained from the difference between these quantities as $C_z=C_z^A+C_z^B+2 C_z^{AB}$.

One technical challenge is to measure the variances ${\rm Var}(\hat S_z^{\sigma})$ at the level of the quantum projection noise. This requires one to repeat the experiment many times in order to reduce finite sampling noise; we typically record eight series of $70$ pictures. In addition, we have performed an exhaustive data analysis by carefully characterizing each fundamental and technical noise source in our detection scheme, using the law of total (co-)variances. We have also employed the Delta method \cite{Oehlert1992} to account for corrections to our variance measurements arising from fluctuations in both the total and sub-ensemble atom numbers (for details, see \cite{AlaouiThesis}).

Results for correlators are shown in Fig.~\ref{fig2}. Note that the atom number is time-dependent due to dipolar relaxation losses at short times. These affect the doubly occupied sites, which exist mostly at the center of the typical wedding-cake distribution \cite{Gemelke2009} obtained in the Mott regime: atoms in these sites are lost typically after 10 ms. These losses have been shown not to substantially affect the spin properties of the shell with unit-filling that survives  \cite{Alaoui2022}. We find that negative correlations $C_z<0$ develop within the whole sample (Fig.~\ref{fig2}a), and recover prior results obtained \textit{without} implementing bipartition \cite{Alaoui2022}: this shows that our bipartition process does not suffer from systematic effects that could arise, e.g. from collisions. In contrast, we find that positive correlations build up  within the subsystems, while the two subsystems become anti-correlated (Fig.~\ref{fig2}b). This measurement of the anisotropy of correlators via bipartition constitutes the main result of this work. We discuss now how this arises from both the anisotropy of the dipolar interactions and the nature of the bipartition geometry.

{\it Thermalization to a negative spin temperature.}
In order to gain physical insight, we first discuss results of our calculations for the thermalized state that is expected at long time for our system, assuming it to be chaotic and undergo quantum thermalization \cite{Lepoutre2019}. We characterize this state by a thermal-like density matrix, which should in principle account for all conserved quantities.
While all powers of $\hat{H}$ and $\hat{S}_z$ are conserved in the dynamics, we only require the thermal state to reproduce the average energy and variance of $\hat{S}_z$. These constitute all the conserved quantities which are quadratic in the spin moments, and which thus most directly influence the two-point correlations of interest. Moreover the variance of $\hat{S}_z$ completely determines the $\hat{S}_z$ distribution in the initial state, since the fluctuations of $\hat{S}_z$ are Gaussian. Hence, we take $\hat\rho \propto \exp(-\beta \hat H - \mu_2 \hat{S}_z^2)$, where $\beta$ is the inverse temperature, and $\mu_2$ a Lagrange multiplier. We have implicitly also included a term $\mu_1 \hat{S_z}$ with $\mu_1=0$, which fixes $\langle \hat{S}_z \rangle=0$.

In the case of a high thermalization temperature, which we expect for our system and initial conditions, we can expand $\hat{\rho}_{\rm th}$ to first order; corresponding expressions for $\beta$ and $\mu_2$ are given in \cite{SuppMat}. Assuming $N_A=N_B=N/2$, we obtain the following expression for thermal correlators:
\begin{eqnarray}
C_{z ,\rm th}^\sigma=-32 N \left( \beta \overline{V_\sigma}+\mu_2 N/4 \right).
\label{Czthermal}
\end{eqnarray}
Subsystems correlators are given by $\overline{V_A}=\overline{V_B}= \sum_{i_A\neq j_A} V_{i_A,j_A}/2N$ , and the intersystem correlator is given by $\overline{V_{AB}}=\sum_{i_A,j_B} V_{i_A,j_B}/2N$. Eq.~(\ref{Czthermal}) shows that $C_{z, \rm th}^{A}=C_{z, \rm th}^{B}=C_{z, \rm th}^{AB}=C_{z, \rm th}/4$ if $\overline{V_A}=\overline{V_{AB}}$. On the other hand, the difference between $C_{z, \rm th}^{A}=C_{z, \rm th}^{B}$ and $C_{z, \rm th}^{AB}$ reveals the anisotropic nature of our system. We numerically evaluate $\overline{V_A}$ and $\overline{V_{AB}}$, which depend on both lattice structure and size \cite{SuppMat}, and infer thermal correlators for a given $B_Q$, as shown in Fig.~\ref{fig2}c. Our measurements, which indicate that $C_{z }^{A}$ and $C_{z }^{B}$ ($C_{z}^{AB}$) evolve towards a positive (negative) value, exclude thermalization at positive temperature as well as  a positive sign of the tensor light-shift term, with a high confidence margin as shown in Fig.~\ref{fig2}c. By contrast, these conclusions cannot be reached when measuring $C_z$ only, as the value of $C_z$ is compatible with both a positive and a negative $B_Q$: bipartition enables us to pinpoint the up-to-now elusive sign of $B_Q$, whose numerical value can be inferred from spectroscopic data only at the kHz level. This demonstrates the utility of bipartite measurements for capturing new fundamental features of quantum thermalization.

Our results thus imply that the spin degrees of freedom thermalize to a high, yet negative temperature ($T\simeq-2.5$ nK). Such thermalization at negative temperature \cite{Kuzemsky2022} has been observed only for external degrees of freedom in prior cold-atom experiments \cite{Braun2013}.
We stress the versatility of our system in which, using the same lattice, thermalization can be tuned to large positive temperature by controlling either the sign of $B_Q$ or the orientation of the magnetic field (see the expression for $\beta$ in \cite{SuppMat}).

{\it Numerical methods.}
We now turn to our numerical modeling of the correlation dynamics.
At short time, exact second-order short-time expansions are obtained for all quantities (see \cite{SuppMat}). We find that no simple relationship exists between short-time and long-time behaviour, as we will discuss below for the microscopic correlations.
We also have developed various numerical simulations to compare with data at intermediate times. First, we have used  calculations based on truncated cumulant expansion (TCE), in which order $n \geq3$ cumulants of quantum fluctuations are set to zero (see \cite{SuppMat}).
Secondly, we have performed simulations based on the GDTWA \cite{Schachenmayer2015,Zhu2019}, which has been successfully used in comparison with magnetic-atom experiments in the recent past \cite{Lepoutre2019,Patscheider2020,Alaoui2022}. Interestingly, we have found that GDTWA appears to overestimate the expected thermalization value predicted by our high-temperature expansion for subsystems correlators (see also \cite{Muleady2023}). In order to improve the GDTWA prediction for correlations, we have developed a cluster-GDTWA approach that combines elements of prior approaches \cite{Zhu2019,Wurtz2018}, in which neighboring pairs of spins are clustered together, so that their quantum correlations are treated exactly, while the correlations between clusters are treated semi-classically. Cluster-GDTWA leads to a marked improvement over standard GDTWA in capturing the long-time thermal predictions \cite{SuppMat}. While we expect further improvement by using larger clusters, we limit our application here to pairs of spin-3 particles owing to the complexity that is exponential in the size of each cluster.

At short-times all three theoretical predictions for the dynamics coincide. After about 7 ms,  differences begin to arise between cluster-GDTWA and TCE, which we interpret as the onset of cumulants with order 3 and higher in the quantum spin fluctuations.
We point out that none of our models include the dipolar losses that affect the doubly-occupied sites at short times, and that apparently do not affect the agreement with measured bipartite correlations.

\begin{figure*}
\centering
\includegraphics[width= 7.0 in]{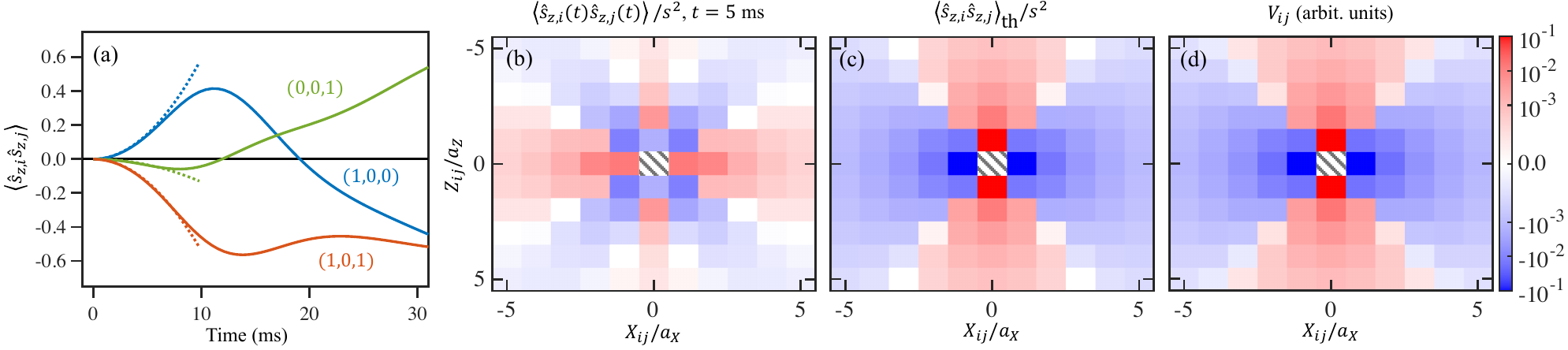}
\caption{\setlength{\baselineskip}{6pt} {\protect\scriptsize Simulations for correlations at the microscopic level. In a given $XZ$ plane, the bipartition separates odd and even sites along the $X$ axis. (a) Dynamics of the $\left<\hat{s}_{z,i}\hat{s}_{z,j}\right>$ correlators corresponding to sites $i$ and $j$ separated by one lattice site along the $X$ dimension (blue), $Z$ dimension (green), or both (orange), denoted by difference vectors $(1,0,0)$, $(0,0,1)$, and $(1,0,1)$ respectively, computed via cluster-GDTWA. Corresponding short-time expansions are also shown (dotted). (b) Off-site correlations at $t=5$ ms computed via a short-time expansion, and normalized by its maximum value $s^2$. We show these correlations as a function of the $X$/$Z$ distances between lattice sites $i$ and $j$, denoted $X_{ij}$/$Z_{ij}$, and normalized by the corresponding lattice spacing $a_X$/$a_Z$. Values of corresponding (c) thermal correlators and (d) interaction matrix elements. Values in (b-d) are shown on a symmetric logarithmic scale to highlight behavior at larger distances.}}
\label{fig3}
\end{figure*}

{\it  Microscopic structure of the correlations.}
We now turn to the dynamics of correlations at the microscopic scale, as revealed by our theoretical approaches. In Fig.~\ref{fig3}, we show our results for the correlation maps in the $XZ$ plane, containing the strongest couplings in our
3D anisotropic lattice; even/odd columns in Fig.~\ref{fig3} correspond to subsystems $A$ and $B$ in Fig.~\ref{fig1}. We compare the cluster-GDTWA calculations and short time expansions for the correlators (Fig.~\ref{fig3}a,b) to their thermal values given in Fig.~\ref{fig3}c.
We find that the next-nearest-neighbor correlations initially grow more rapidly than the nearest-neighbor correlations, which can be explained by our second-order short time expansion, see \cite{SuppMat}.
Moreover, nearest-neighbor correlations change in sign along the time evolution, as shown in Fig.~3a.
We find that this peculiar behavior is partly a consequence of the negative sign of $B_Q$. Indeed, our short-time expansion shows that for sufficiently large $B_Q$, $\left<\hat{s}_z^i\hat{s}_z^j\right> \propto \mathrm{sgn} (B_Q) V_{i,j}$; thus, at short time, the sign of the correlator and of the coupling between two given sites can be opposite for $B_Q<0$. In the thermal state, the correlation map is instead identical to the interactions map (Fig.~\ref{fig3}c-d) up to a global shift, i.e. $V_{i,j} =  -16\beta\left<\hat{s}_z^i\hat{s}_z^j\right> + {\rm const}$ \cite{SuppMat}. Our microscopic analysis thus demonstrates that the sign inversion of  correlators at short distance is deeply linked to thermalization at high negative temperature, dictated by the negative value of $B_Q$.
Interestingly, the sign of the subsystem correlator $C_z^{AB}$ remains negative throughout the evolution.
At short times $C_z^{AB}$ is dominated by the negative value of the correlations at separation $(X,Y,Z) = (1,0,1)$ (Fig.~\ref{fig3}a; at longer times the correlations at separation $(1,0,0)$ turn negative and further contribute to the negative value of  $C_z^{AB}$.

{\it  Conclusion.}
In this work we have experimentally implemented  a bipartition of magnetic atoms in a 3D optical lattice prepared in  a Mott insulator state. We have used this scheme to  track short-range  spin correlations without the need for individual addressing. The observed structure of correlations reveals the anisotropic nature of our dipolar spin system. In this work we focused on correlations of a single spin projection axis; extending this analysis to orthogonal spin components would enable entanglement certification \cite{Giovannetti2003}.

Comparing our results to various numerical methods led to several additional insights. The predictions of a thermal ensemble with fixed magnetization distribution, when matched to our late-time experimental results, allowed us to unambiguously fix the Hamiltonian parameters, and to conclude that thermalization takes place at a high, negative temperature. The observed evolution of the correlation map is highly non-trivial, with correlations changing sign between the short-time evolution and the long-time one.
Our work demonstrates that magnetic-atom quantum simulators can act as an efficient test bed for state-of-the-art models of out-of-equilibrium quantum dynamics, and exhibit highly non-trivial quantum thermalization, due to the power-law and anisotropic nature of dipolar interactions \cite{Sugimoto2022,Levin2014}.

\vspace{1cm}

\begin{acknowledgments}
We  acknowledge careful review of this manuscript and  useful  comments from Joanna W. Lis and Sanaa Agarwal.
The Villetaneuse and Lyon groups acknowledge financial support from CNRS. The Villetaneuse group acknowledges financial support from Agence Nationale de la Recherche (project EELS - ANR-18-CE47-0004), QuantERA ERA-NET (MAQS project). S.R.M. is supported by the NSF QLCI grant OMA-2120757. A.M.R. is supported by the AFOSR grant FA9550-18-1-0319, AFOSR MURI, the ARO single investigator award W911NF-19-1-0210,   the NSF JILA-PFC PHY-2317149 grants, and by NIST.
\end{acknowledgments}

\vspace{10cm}
\section{Supplemental material}

\subsection{Additional data}
\begin{figure}
\centering
\includegraphics[width= 3.3 in]{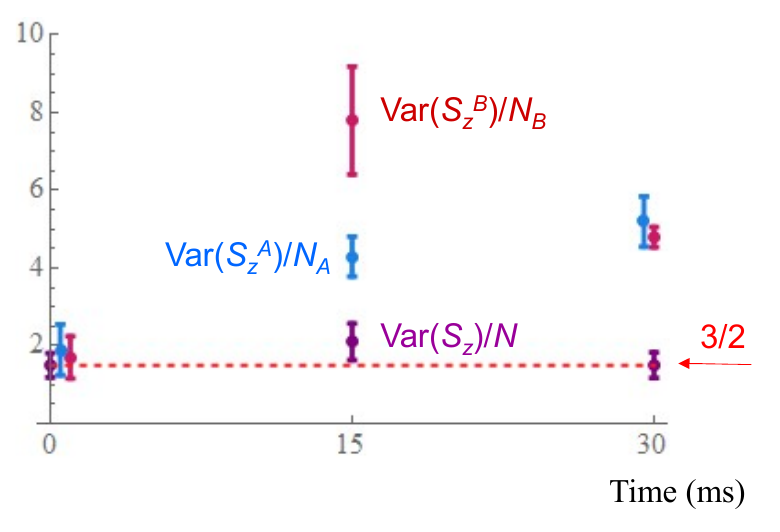}
\caption{\setlength{\baselineskip}{6pt} {\protect\scriptsize Experimental results for the variance of the magnetization. Variances are normalized to the relevant atom number. Error bars correspond to two statistical standard errors. Data at $t=0$ and $t=30$ ms are offset for clarity. }}
\label{FigVar}
\end{figure}

We present in Fig \ref{FigVar} our results regarding magnetization variances ${\rm{Var}}(\hat S_z)$ (whole system), and ${\rm{Var}}(\hat S_z^A)$, ${\rm{Var}}(\hat S_z^B)$ (subsystems). While the variance of the total magnetization remains close to $3 N/2$ at all times,  magnetization fluctuations of the two spin sub-ensembles dynamically become larger than the one of an uncorrelated sample $3 N_{A,B}/2$. This is in agreement with the growth of a negative covariance $C_z^{AB}={\rm Cov}(\hat{S}_z^A,\hat{S}_z^B)$, see main article.

\subsection{Description of the lattice and bipartition}
\begin{figure}
\centering
\includegraphics[width= 3.3 in]{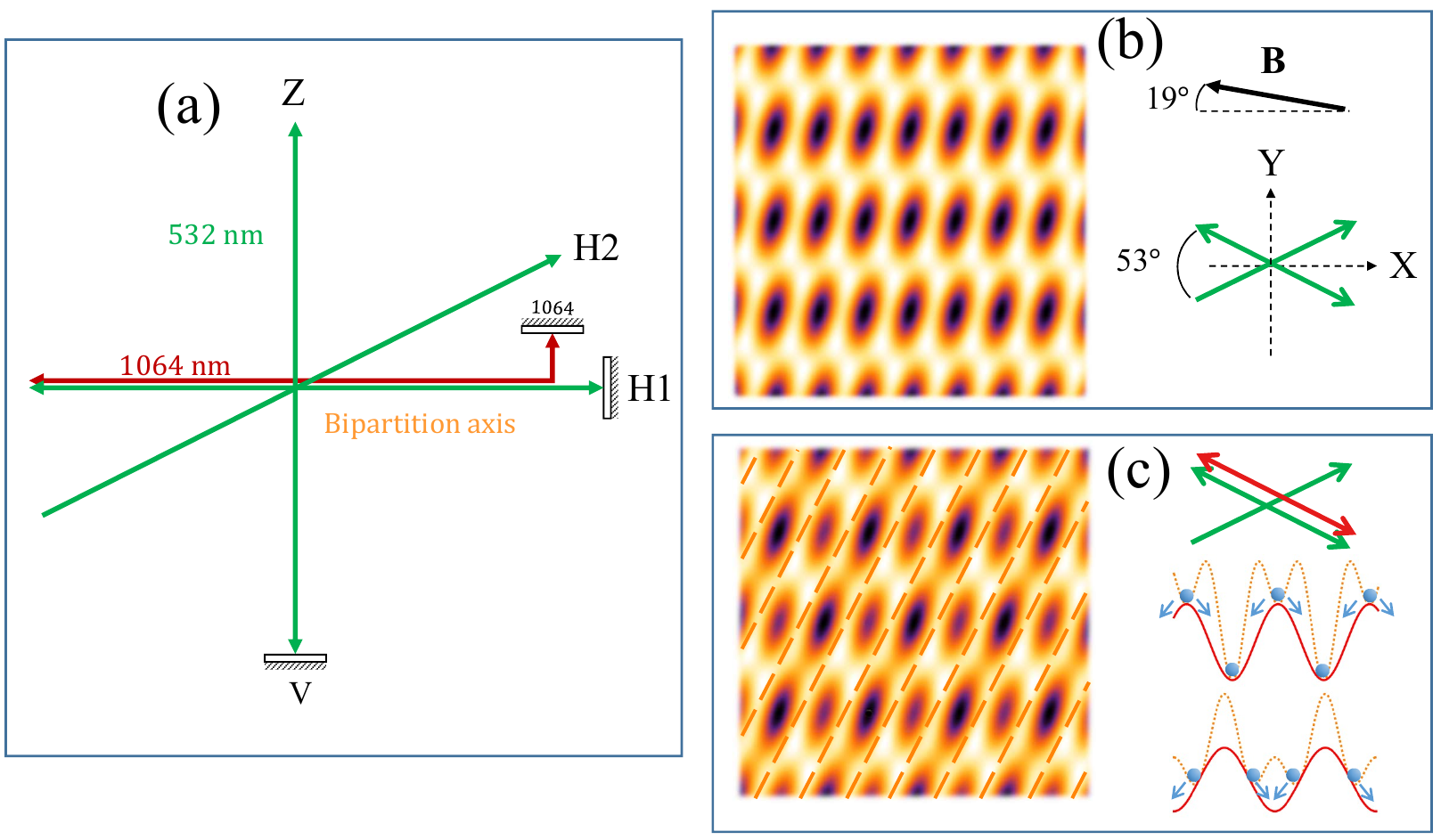}
\caption{\setlength{\baselineskip}{6pt} {\protect\scriptsize (a) Implementation of the 3D $532$ nm lattice in which atoms interact. (b) Lattice potential in the horizontal plane. (c) Lattice potential in the horizontal plane when an IR lattice is added, at the end of spin dynamics. The potential depends on the phase between lasers, the bottom potential is well adapted to obtain bipartition after time of flight, while the top one leads to three clouds.}}
\label{FigLattice}
\end{figure}

Here, we provide information on the 3D tetragonal lattice in which the dipolar dynamics takes place, and describe the implemented bipartition. The 3D lattice is created by 5 lasers at $532$ nm. In the horizontal plane, three laser beams interfere, generating an approximately rectangular lattice with lattice constants $a_X = 300$ nm and $a_Y = 600$ nm along $X$ and $Y$, respectively. A retro-reflected laser beam creates an independent interference pattern along the vertical axis with lattice constant $a_Z = 270$ nm (Fig \ref{FigLattice}a. The potential pattern in the horizontal plane is shown in Fig \ref{FigLattice}b; axes X and Y are the eigen-axes, and correspond to bisectors between the two horizontal laser beams. The magnetic field lies in the horizontal plane, at an angle of $19$ from X. The addition of an IR laser at $1064$ nm modifies the optical potential, and generates two sublattices (Fig \ref{FigLattice}c). The total potential along X at the position of the atoms depends on the relative phase of the $532$ nm and $1064$ nm lasers, and extreme cases are shown. After moving in the pure IR potential only, atoms belonging to two different sublattices acquire two well defined opposite velocities in the favourable case, hence a good spatial separation after time-of-flight and a successful bipartition. In the other case, one obtains three different clouds. The transition between the two cases is observed on a $\simeq$ 20 mn period due to relative frequency drifts of the two lasers, when no frequency correction is applied.

\subsection{Short time expansions of correlators}
Here, we derive the short-time expansion for the spin correlators shown in the main text. For sufficiently small $t$, we may expand
\begin{align}
    \hat{s}_{z,i}(t) = \hat{s}_{z,i} + \frac{d\hat{s}_{z,i}}{dt}\bigg\lvert_{t=0} t + \frac{1}{2}\frac{d^2\hat{s}_{z,i}}{dt^2}\bigg\lvert_{t=0} t^2+ O(t^3),
\end{align}
where the relevant derivatives can be directly obtained from the Heisenberg equations of motion. Also, for the initial spin-polarized state, the expectation values at $t=0$ can also be obtained for any string of spin operators. From this, we directly obtain
{\small
\begin{align}
    \braket{\hat{s}_{z,i}(t)\hat{s}_{z,j}(t)}_{i\neq j} & \approx \frac{t^2s^2}{16} \times \\
    & \left[6s\left(\sum_k V_{ik}V_{kj}  - 2V_{ij} \overline{V}\right) + 3V_{ij}^2 + 8\left(s-\frac{1}{2}\right)V_{ij}B_Q\right] \nonumber
\end{align}
}
where, as before, $\overline{V^n} = \frac{1}{2N}\sum_{i\neq j} V_{ij}^n$.

Summing this correlator for various $i$, $j$, we have
{\small
\begin{align}
    C_z(t) &= \sum_{i\neq j}^N \braket{\hat{s}_{z,i}(t)\hat{s}_{z,j}(t)} \approx \frac{-t^2 s^3 N}{4}\left(1-\frac{1}{2s}\right)\left(3\overline{V^2} - 4\overline{V}B_Q \right)\\
    C_z^A(t) &= \sum_{i_A\neq j_A}^N \braket{\hat{s}_{z,i}(t)\hat{s}_{z,j}(t)} \approx \frac{-t^2 s^3 N}{4} \times\\
 &   \left[ 3\overline{V_{AB}^2} + 12\overline{V_{AB}}\left(\overline{V_{A}} - \overline{V_{AB}}\right) + \left(1 - \frac{1}{2s}\right)\left(3\overline{V^2_A} - 4\overline{V_A}B_Q \right)\right] \nonumber \\
    C_z^{AB}(t) &= \sum_{i_A, j_B}^N \braket{\hat{s}_{z,i}(t)\hat{s}_{z,j}(t)} \approx\frac{t^2 s^3 N}{4} \times \\
   & ( 3 \overline{V_{AB}^2} + 12\overline{V_{AB}}\left(\overline{V_{A}} - \overline{V_{AB}}\right) \nonumber \\
   & - \left(1 - \frac{1}{2s}\right)\left(3\overline{V^2_{AB}} - 4\overline{V_{AB}}B_Q \right)) \nonumber
\end{align}
}
For these latter expressions, we have assumed translational invariance of $V_{ij}$, so that we may define $V(\mathbf{r}_{ij}) \equiv V_{ij}$ where $\mathbf{r}_{ij} \equiv \mathbf{r}_i - \mathbf{r}_j$. We assume the property that $V(\mathbf{r}_{ij}) = V(-\mathbf{r}_{ij})$, as is the case for dipolar interactions. Then,
{\small
\begin{gather}
    \sum_{i\neq j} V_{ij}^n = N\sum_{\mathbf{r}\neq \mathbf{0}} V(\mathbf{r})^n = 2N\overline{V^n} \\
    \sum_{k\neq i,j} V_{ik}V_{kj} = \sum_{\mathbf{r}\neq \mathbf{0}} V(\mathbf{r})V(\mathbf{r} + \mathbf{r}_{ij})\\
     \sum_{i\neq j\neq k} V_{ik}V_{kj} = \sum_{i, j,\mathbf{r}\neq \mathbf{0}}V(\mathbf{r})V(\mathbf{r} + \mathbf{r}_{ij}) = 4N\overline{V}^2.
\end{gather}
}
For the sum over sublattice coordinates, we likewise may obtain
\begin{align}
\sum_{i_A\neq j_A\neq k} V_{i_Ak}V_{kj_A} = 8N\left(\overline{V_A}^2 + \overline{V_{AB}}^2\right) \\
\sum_{i_A\neq i_B\neq k} V_{i_Ak}V_{ki_B} = 16N\overline{V_A}\,\overline{V_{AB}}.
\end{align}

\subsection{High temperature expansion}
The expected steady-state values for these correlators can be obtained by assuming the system relaxes to high-temperature thermal state. In order to properly account for the distribution of the initial state over various $\hat{S}_z$ symmetry sectors, we include the variance of the collective spin ($\hat{S}_z^2$ for $\langle\hat{S}_z\rangle = 0$) as a conserved quantity in our thermal ensemble,
\begin{gather}
    \hat{\rho}_{\mathrm{th}} = \frac{1}{Z}\exp\left\{-\beta\hat{H} - \mu_1\hat{S}_z - \mu_2\hat{S}_z^2\right\}\label{eq:rho_therm_Cr} \\
     Z = \mathrm{Tr}\left[\exp\left\{-\beta\hat{H} - \mu_1\hat{S}_z - \mu_2\hat{S}_z^2\right\}\right] \nonumber
\end{gather}
for inverse temperature $\beta$ and potentials $\mu_1$ and $\mu_2$. Observables in this state are then given by $\langle \hat{O}\rangle_{\mathrm{th}} = \mathrm{Tr}[\hat{O}\hat{\rho}_{\mathrm{th}}]$.

Assuming the high-temperature limit, and that $\mu_1$ and $\mu_2$ are sufficiently small, we can expand this to first order in these parameters. This results in the following expression for the correlators
{\small
\begin{align}
    \langle\hat{s}_{z,i}\hat{s}_{z,j}\rangle_{\mathrm{th}} &\approx -16(1-\delta_{ij})(\beta V_{ij} + 2\mu_2) + \delta_{ij}\left[4 - 12(\beta B_Q + \mu_2)\right]
\end{align}
}
with temperature and potential
{\small
\begin{gather}
\beta = \frac{5 B_Q + 9 \overline{V}}{24 B_Q^2 + 48 \overline{V^2}},\qquad \mu_2 = \frac{1}{192 N}\left[\frac{30 \overline{V^2} - (67 B_Q  + 72\overline{V})\overline{V}}{B_Q^2 + 2\overline{V^2}}\right] \label{eq:beta_mu2_pi_2}
\end{gather}
}
as $N\rightarrow \infty$. Here, we have the interaction moments $\overline{V^n} = \sum_{i\neq j} V_{ij}^n/2N$, and assume these are independent of $N$. Previous work has reported values of $\overline{V} \approx -0.57$ Hz, and $\sqrt{\overline{V^2}} \approx 4.35$ Hz; we also examine this below.

Assuming this thermal state represents the steady-state of the dynamics, then in the large $N$ limit we expect
\begin{align}
    C_z(t\rightarrow \infty)/N &= \frac{1}{N}\sum_{i\neq j} \mathrm{Cov}(\hat{s}_i^z,\hat{s}_j^z)_{\mathrm{th}} = -32\left(\beta \overline{V} + N\mu_2\right),\label{eq:Cz_therm1}
\end{align}
or equivalently,
\begin{align}
    C_z(t\rightarrow\infty)/N &= 3/2 - \frac{1}{N}\sum_i\mathrm{Var}( \hat{s}_i^z)_{\mathrm{th}} = -5/2 + 12 \beta B_Q \label{eq:Cz_therm2}
\end{align}
owing to the enforced conservation of $\langle\hat{S}_z^2\rangle$. We note that without consideration of $\mu_2$, these expressions are \emph{not} equivalent, with Eq.~\eqref{eq:Cz_therm1} and Eq.~\eqref{eq:Cz_therm2} predicting steady-state values of $\approx -0.37$ and $\approx -1.27$, respectively, when $B_Q = -5.1$ Hz. However, with the inclusion of $\mu_2$, both expressions consistently yield the same value of $\approx -1.27$.

For the sublattice correlators we have
\begin{align}
    C_z^{A}(t\rightarrow\infty)/N = -32\left(\beta\overline{V_{A}} + \mu_2 N/4\right) \\ C_z^{AB}(t\rightarrow\infty)/N = -32\left(\beta\overline{V_{AB}} + \mu_2 N/4\right)
\end{align}
for $\overline{V}_A = \sum_{i_A\neq j_A} V_{i_Aj_A}/2N$, $\overline{V}_{AB} = \sum_{i_A, j_B} V_{i_Aj_B}/2N$, which depend on the relevant lattice geometry.

\subsection{Computing interaction moments}
\begin{figure}
    \centering
    \includegraphics[width=0.8\linewidth]{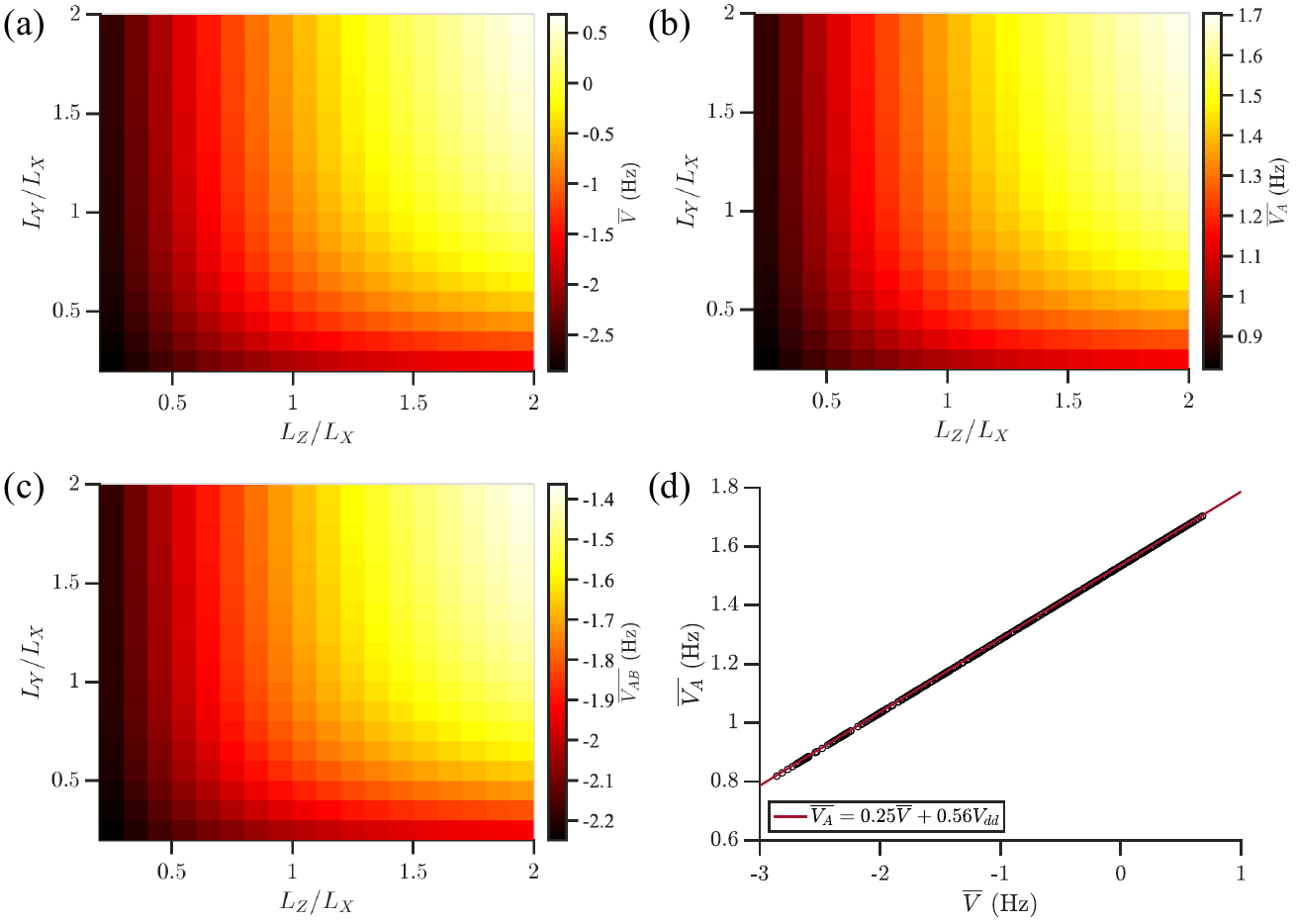}
    \caption{Values of dipolar moments (a) $\overline{V}$, (b) $\overline{V_A}$, and (c) $\overline{V_{AB}}$ for varying lattice sizes, with $L_X = 100$. (d) Scatter plot of $\overline{V_A}$ versus $\overline{V}$ for all system sizes considered, and corresponding linear fit $\overline{V_A} = 0.25\overline{V} + 0.56V_{dd}$. We note that $\sqrt{\overline{V^2}}$ remains essentially fixed at $4.35$ Hz for all lattice sizes considered.}
    \label{fig:fig_dipolarIntegral}
    \label{fig:fig_dipolarFit}
\end{figure}

We compute the values of the interaction moments with the lattice geometry relevant for our lattice geometry, and corresponding moments on the considered sublattice geometry. In particular, we are interested in determining the values of $\overline{V} = (1/2N)\sum_{i\neq j} V_{ij}$ and $\overline{V^2} = (1/2N)\sum_{i\neq j} V_{ij}^2$, as well as the corresponding sublattice sums $\overline{V_A} = (1/2N)\sum_{i_A\neq j_A} V_{i_Aj_A}$, $\overline{V_{AB}} = (1/2N)\sum_{i_A, j_B} V_{i_Aj_B}$. In Fig.~\ref{fig:fig_dipolarIntegral}, we show the numerically computed values on a lattice with $L_X = 100$, and varying values for $L_Y$ and $L_Z$, where $L_X$, $L_Y$, and $L_Z$ denote the number of lattice sites along each dimension. Although the values of these sums vary heavily depending on the precise system dimensions being considered, the corresponding value of $\overline{V_A}$ is found to be well-captured via the relation $(\overline{V}/V_{dd}) = 0.25(\overline{V_A}/V_{dd}) + 0.56$ also shown in Fig.~\ref{fig:fig_dipolarFit}, and is thus uniquely determined by the value of $\overline{V}$ (and so $\overline{V_B}$ and $\overline{V_{AB}}$ are also uniquely determined from the relationships $\overline{V_{A}} = \overline{V_{B}}$ and $\overline{V} = 2\overline{V_A} + 2\overline{V_{AB}}$).

Throughout, as in previous work~\cite{Lepoutre2019}, we take $\overline{V} = -0.57$ Hz and $\sqrt{\overline{V^2}} = 4.35$ Hz, which are the values obtained near $L_X = L_Y = L_Z$. From our scaling relation, as well as relations between these interaction moments, this directly leads to $\overline{V_A} = 1.40$ Hz and $\overline{V_{AB}} = -1.69$ Hz.

\subsection{Truncated cumulant expansion}

The truncated cumulant expansion (TCE) is a standard technique to close the hierarchy of differential equations governing the evolution of observables in interacting many-body systems
\cite{Rudiger1990,Leymann2014,Plankensteiner2022,Colussi2020,Verstraelen2023}.  Cumulants $\langle ... \rangle_c$ of products of local observables $A_i$ for a lattice system are generically defined recursively as
\begin{align}
& \langle A_i \rangle  =  \langle A_i \rangle_c \\
&\langle  A_i A_j \rangle  =  \langle A_i A_j \rangle_c + \langle A_i \rangle_c \langle A_j \rangle_c \\
&\langle  A_i A_j A_k \rangle  =  \langle A_i A_j  A_k\rangle_c + \langle A_i A_j  \rangle_c \langle A_k\rangle_c  \\
&+ \langle A_i A_k  \rangle_c \langle A_j\rangle_c + \langle A_j A_k  \rangle_c \langle A_i\rangle_c +   \langle A_i \rangle_c \langle A_j \rangle_c \langle A_k \rangle_c
\nonumber \\
& (...) \nonumber~.
\end{align}
Assuming that all cumulants beyond a given order vanish allows one to express any product of local observables $\langle A_{i_1}  A_{i_2} ... \rangle$ in terms of the finite cumulants of lower order.

In a lattice spin system with spins of length $s$ we can decompose any local operator in terms of the elementary operators $T_i^{m_s,m_s'} = |i;m_s\rangle \langle i; m_s'|$ with $m_s, m_s' = -s, ...., s$. The second-order TCE applied to these operators consists \cite{Trifa2023} in writing the Heisenberg equation of motion for the single-site averages $\langle T_i^{m_s,m_s'}  \rangle$ and two-site ones $\langle T_i^{m_s,m_s'} T_j^{n_s,n_s'}  \rangle$. These equations of motion contain generically three-site averages $\langle T_i^{m_s,m_s'} T_j^{n_s,n_s'} T_k^{p_s,p_s'} \rangle$, whose expression can nonetheless be written in terms of the single-site and two-site terms by postulating the vanishing of the corresponding cumulant $\langle T_i^{m_s,m_s'} T_j^{n_s,n_s'} T_k^{p_s,p_s'} \rangle_c=0$.

The logic behind the TCE approach to non-equilibrium quantum dynamics is that cumulants of order higher than one are all vanishing in the initial coherent spin state, because of its factorized nature. Hence we can suppose that higher-order cumulants will appear progressively in the evolution, starting from second order; and that, for sufficiently short times, one can safely ignore cumulants beyond a given order. Neglecting cumulants of order 2 and higher is equivalent to a simple mean-field approximation -- which cannot account for the buildup of correlations, at the core of this work. Including cumulants of order 2, but neglecting those of order 3, is instead the minimal strategy to describe the onset of correlations at a minimal numerical cost. A beneficial aspect associated with this approach is that both the energy as well as the variance of the $S^z$ magnetization -- namely the two main conserved quantity in the dynamics -- can be expressed as second-order cumulants, and they are therefore explicitly conserved in the equations of motion. It is also important to stress that a second-order TCE does not simply coincide with an assumption of Gaussian statistics for the spin fluctuations, because e.g. the physics of both individual spins and pairs of spins is treated exactly in this approach -- since we work with the bases $T_i^{m_s,m_s'} $ and  $T_i^{m_s,m_s'} T_j^{n_s,n_s'}$ for all single-spin and two-spin observables. Hence highly non-Gaussian fluctuations at the level of single spins or pairs of spins can be described via the TCE we developed.

The TCE strategy is bound to break down at intermediate times, as the spreading of quantum correlations in the system leads to irreducible correlations among three sites, four sites, etc. resulting in cumulants to all order. It remains nonetheless rather valuable to compare the predictions from second-order TCE with those of the experiments and of the GDTWA approach, as the departure between the TCE results and the more accurate experimental or theoretical ones signals the onset of increasingly complex quantum correlations in the system.

\subsection{GDTWA simulations}
Here we outline GDTWA, and our cluster extension thereof. For a lattice of spin-$s$ particles, the relevant Hilbert space is given by $\mathcal{H} = \otimes \mathcal{H}_i$, where $\mathcal{H}_i$ is the $(2s+1)$-dimensional Hilbert space for the spin at site indexed by $i$.

To accurately account for the population of individual populations of the Zeeman levels we solve for the dynamics of each local set of $SU(2s+1)$ generators, which we take to be the set of generalized Gell-Mann matrices $\hat{\Lambda}_{\mu,i}$ for $\mu = 1,...,[(2s+1)^2-1]$, see e.g.~\cite{Zhu2019}. These matrices are traceless, as well as trace orthonormal, i.e. $\mathrm{Tr}\left[\hat{\Lambda}_{\mu,i}\hat{\Lambda}_{\nu,j}\right] = \delta_{ij}\delta_{\mu\nu}$, and thus constitute a complete operator basis for each local Hilbert space. The Heisenberg equations of motion for these operators then take the generic form
\begin{align}
    \frac{d\hat{\Lambda}_{\mu,i}}{dt} = i \mathcal{M}_{\mu\nu,i} \hat{\Lambda}_{\nu,i} + i\mathcal{C}_{\mu\nu\rho,ij} \hat{\Lambda}_{\nu,i}\hat{\Lambda}_{\rho,j}\label{eq:HeisenbergEOM}
\end{align}
for
\begin{align}
    M_{\mu\nu,i} = \mathrm{Tr}\left[\left[\hat{H}_i,\hat{\Lambda}_{\mu,i}\right]\hat{\Lambda}_{\nu,i}\right] \\ C_{\mu\nu\rho,ij} = \mathrm{Tr}\left[\left[\hat{H}_{ij},\hat{\Lambda}_{\mu,i}\right]\hat{\Lambda}_{\nu,i}\hat{\Lambda}_{\rho,j}\right],
\end{align}
where tensor products of operators on different sites is implied, and we have assumed our Hamiltonian may be written as $\hat{H} = \sum_i \hat{H}_i + \sum_{i\neq j} \hat{H}_{ij}$.

One can use Eq.~\eqref{eq:HeisenbergEOM} as the basis for a mean field solution to the dynamics by directly replacing each instance of $\hat{\Lambda}_{\mu,i}$ in the expression by its mean value, $\lambda_{\mu,i} = \braket{\hat{\Lambda}_{\mu,i}}$, i.e.
\begin{align}
    \frac{d\lambda_{\mu,i} }{dt} \approx i \mathcal{M}_{\mu\nu,i}
    \lambda_{\nu,i} + i\mathcal{C}_{\mu\nu\rho,ij} \lambda_{\nu,i} \lambda_{\rho,j} \label{eq:MFEOM}
\end{align}
This formally involves the expansion and truncation of the equations of motion to second order in $\hbar$. However, this neglects the role of quantum fluctuations, which play a critical role in the dynamics, see e.g. \cite{Lepoutre2019}. In particular, for the secular dipolar Hamiltonian that drives the dynamics (see Main Article) and an initial uncorrelated spin state polarized along x, the corresponding mean-field solution exhibits no dynamics.

To take into account the role of quantum fluctuations in the dynamics, we utilize a generalization of the discrete truncated Wigner approximation, see \cite{Lepoutre2019,Zhu2019,Alaoui2022}. We first determine a joint probability distribution of measurement outcomes for each basis operator $\hat{\Lambda}_{\mu}$ whose moments correctly reproduce the expectation values all symmetric-ordered observables. We can then solve for the quantum dynamics by evolving this distribution under some suitable equations of motion~\cite{Polkovnikov2011}, which we approximate via Eq.~\eqref{eq:MFEOM}. This retains the role of quantum fluctuations in the system in the form of a semiclassical noise source. Furthermore, in the case that the initial joint probability distribution is non-negative, we can efficiently solve for the dynamics by Monte Carlo sampling initial values for $\lambda_{\mu,i}$, evolving each set of initial values to obtain an ensemble of dynamical ``trajectories'', and then averaging moments of this ensemble to obtain the dynamics of various observables.

For an initial uncorrelated state $\ket{\psi_0}$ with density operator $\hat{\rho}_0 = \ket{\psi_0}\bra{\psi_0}$, we form such an initial distribution of values for $\lambda_{\mu,i}$ by sampling the eigenvalues, or allowed measurement outcomes, of $\hat{\Lambda}_{\mu,i}$. Letting $a_{\mu,i}$ and $\hat{P}_{a_{\mu,i}}$ denote the eigenvalues and corresponding eigenspace projectors of $\hat{\Lambda}_{\mu,i}$, so that $\hat{\Lambda}_{\mu,i} = \sum_{a_{\mu,i}} a_{\mu,i}\hat{P}_{a_{\mu,i}}$, then we sample $a_{\mu,i}$ for the initial value of $\lambda_{\mu,i}$ with probability
\begin{align}
P(a_{\mu,i}) = \mathrm{Tr}\left[\hat{\rho}_0 \hat{P}_{a_{\mu,i}}\right]
\end{align}
and we can likewise independently sample for each site and basis operator. Solving Eq.~\eqref{eq:MFEOM} for each initial set of sampled values, we can then obtain expectation values and symmetrized correlators via
\begin{gather}
    \braket{\hat{\Lambda}_{\mu,i}(t)} \approx \overline{\lambda_{\mu,i}(t)},\\
    \braket{\hat{\Lambda}_{\mu,i}(t)\hat{\Lambda}_{\nu,j}(t) + \hat{\Lambda}_{\nu,j}(t)\hat{\Lambda}_{\mu,i}(t)} \approx 2\overline{\lambda_{\mu,i}(t)\lambda_{\nu,j}(t)} \quad(i\neq j),
    \label{eq:GDTWA_corr}
\end{gather}
where $\overline{\,\cdot\,}$ denotes averaging over the ensemble of sampled trajectories. For expectations of operator products on the same site, the operator product should first be linearized as a sum of basis operators for improved accuracy.

\subsection{Cluster-GDTWA}
GDTWA has previously been shown to accurately reproduce single-particle observables for this system, and can thus be used to estimate collective correlators that may be reduced to single-particle observables. For instance, $C_z = \sum_{i\neq j} \braket{\hat{s}_{z,i}\hat{s}_{z,j}}$ may equivalently be computed via $C_z = 3N/2 - \sum_{i,m_s} m_s^2 P_{m_s}$ owing to the total conservation of $\hat{S}_z$ expected in the dynamics, where $m_s$ denotes the Zeeman sublevel magnetization, which has total occupation $P_{m_s}$. This constitutes a fundamental ambiguity in the definition of the correlators in GDTWA. For instance, for $C_z/N$, we can choose to compute either
\begin{align}
    C_z(t)/N \approx \frac{3}{2} - \frac{1}{N}\sum_{i,m_s} m_s^2 \overline{n_{m_s,i}(t)},\label{eq:GDTWA_corr1}
\end{align}
or
\begin{align}
    C_z(t)/N \approx \frac{1}{N}\sum_{i\neq j} \overline{s_{z,i}(t)s_{z,j}(t)},\label{eq:GDTWA_corr2}
\end{align}
where $s_{z,i}$ and $n_{m_s,i}$ are the classical variables corresponding to the total spin and Zeeman sublevel population operators, $\hat{s}_{z,i}$ and $\hat{n}_{m_s,i}$, respectively. Eq.~\eqref{eq:GDTWA_corr1} takes advantage of the total $\hat{S}_z$ variance conservation to compute $C_z(t)$ in terms of only single-body observables, whereas Eq.~\eqref{eq:GDTWA_corr2} fundamentally relies on computing correlators between the sites. While Eq.~\eqref{eq:GDTWA_corr1} accurately captures the expected relaxation, we find that Eq.~\eqref{eq:GDTWA_corr2} provides relatively incorrect results. Furthermore, for the calculation of $C_{z}^A$, $C_{z}^B$, and $C_z^{AB}$, there is no such conservation law that enables us to reduce these expressions to linear sums of our phase space variables.

To improve the accuracy of our calculations, we form clusters of neighboring pairs of spins along the $X$ dimension, which pairs spins in the $A$ and $B$ subsystems. A similar scheme has been previously employed for $s=1/2$ systems using a continuous phase space sampling scheme; here, we extend these ideas to more general spin-$s$ with the discrete sampling scheme outlined in the prior section. For $s=3$, each such pair of spins is then treated as a local $(2s+1)^2 = 49$-dimensional Hilbert space for our applying our GDTWA calculations. We can thus solve for the GDTWA dynamics of the $SU(49)$ generators, and the correlator between these two spins may be decomposed as a linear sum of these generators. The advantage of this approach is that the internal dynamics of each cluster are treated exactly, while inter-cluster correlations are still treated approximately through GDTWA. However, the complexity also drastically increases, as we must simulate the dynamics of $(N/2) \times (2s+1)^4$ variables, rather than the $N \times (2s+1)^2$ variables in ordinary GDTWA.

For our simulations, we utilize this pair-clustering scheme to simulate the dynamics on a $4\times 4 \times 4$ lattice with periodic boundary conditions, and utilize $5\times 10^3$ trajectories.
\vspace{0.5cm}

\bibliography{biblioBipartitionArXiv}

\end{document}